# A Review of Space Tribology Experiments in Low Earth Orbit: Challenges and Opportunities

H.D.S. Amaradasa, I. Sherrington, N. Renevier, M. Bernabei, A. Ghanbari

Jost Institute for Tribotechnology, University of Central Lancashire, Preston PR1 2HE, UK

## ABSTRACT

Operating mechanical devices in low earth orbit (LEO) environment presents unique challenges due to adverse effects of the LEO environment on lubricants and materials in tribo-mechanisms. These challenges include corrosion due to atomic oxygen, molecular degradation of materials and fluids due to radiation, temperature extremes influencing lubricant viscosity, and rapid evaporative loss of fluids in vacuum conditions. Therefore, lubricants for mechanisms and components such as bearings and gears for spacecraft should be tested extensively in both air and vacuum to ensure their continuous and accurate function. Literature on ground based tribo-testing is extensive and well-established. However, tribological investigations conducted in LEO are much fewer in number. The purpose of this paper is to draw together details of tribology experiments of this type, to try to clarify their purpose and value. This review presents these studies according to a thematic categorization of the mechanisms involved.

## 1. INTRODUCTION

The space industry is rapidly expanding with both organizations and private companies investing substantially in technologies for use in space. This includes ambitious projects for exploiting resources in space, launching constellations of earth-orbiting satellites and exploring beyond Earth. According to Euroconsult forecasting, the "new space market is expected to grow from \$12.6 billion to \$42.8 billion in the next decade (2019-2028) [1]. However, it is widely acknowledged that space missions are inherently complex and costly undertakings. Consequently, component and system failures can have a significant impact on the financial stability and future prospects of the organizations involved in operating them as well as delaying significant scientific outcomes and limiting commercial success.

The reliable functioning of spacecraft mechanisms is a vital aspect of any space mission to facilitate the dependable deployment, articulation, pointing, and separation of various subsystems and payloads. A single spacecraft may contain numerous mechanisms, including solar arrays, antennas, thrusters, reaction wheels, hinges, latches, and valves [2-5]. All these moving mechanisms necessitate lubrication to function and perform for the expected duration of the mission.

In the early decades of the space program, spacecraft mission lifetimes were limited by the failure of subsidiary electronics such as batteries and computers. As a result, the expected lifetime of space mechanisms was minimal [5]. However, over the past two decades, electronic systems have been substantially improved and tribological systems are now one of the limiting factors in spacecraft performance. In recent missions, a substantial number of spacecraft mechanism failures can be attributed to tribological-related failures [6]. Hence, it is imperative to understand the tribological performance of spacecraft mechanisms with the intention of preventing or reducing failures to ensure the success of future space missions. A substantial amount of open literature is dedicated to the analysis of fast rolling mechanisms, particularly in association with bearing failures. Thus, this paper mainly discusses studies related to rolling mechanisms. However, it is noted that there are other types of space mechanisms, which involve linear, reciprocating, sliding, and oscillating motion, which may operate at slower speeds or with one-time movements. Conducting tribological tests that accurately replicate space conditions within a laboratory presents a significant challenge. The difficulty lies in the fact that it is not a trivial task to simulate a space environment which simultaneously reproduces the conditions of atomic oxygen, radiation, vacuum, temperature, and microgravity found in LEO [2]. Consequently, tribological data is often obtained from tests conducted in laboratories under each individual condition with a further useful, but limited, level of understanding being derived from experiments and operation of mechanisms in LEO [7]. As demands on space engineering continue to grow, more innovative approaches to conducting accurate tribological tests are required. Details and results from tribological tests conducted in a terrestrial setting are well documented in the open literature. However, this paper presents a review of tribological experiments undertaken in LEO while reflecting on their degree of realism. The paper also briefly considers potential future developments relating to tribology in space.

## 2. LOW EARTH ORBIT

Low earth orbit refers to an orbit around Earth at an altitude that falls within the lower range of possible orbits, typically between 160 and 2,000 km above Earth [8]. Spacecraft in LEO generally have an orbital period of 128 minutes or less [9]. Approximately 85% of active satellites operate in LEO [10], including the International Space Station (ISS), the Hubble Space Telescope (HST), satellites belonging to the Global System (GPS) and recently launched Starlink Internet satellites. Owing to the proximity to Earth, spacecraft in this region can benefit from lower launch costs, shorter communication delays, and easier access for maintenance and repair. However, spacecraft in LEO experience a dynamic environment with high levels of atomic oxygen (AO), ultraviolet (UV) radiation, ionizing radiation, vacuum and thermal extremes. The constituents of natural orbital environment vary with position, local time, season and solar activity [11]. Due to the proliferation of satellites in LEO, conducting tribological tests on space mechanisms in LEO has become more economically feasible.

## 3. EFFECTS OF THE LEO ENVIRONMENT ON TRIBOMECHANISMS

### 3.1. Atomic oxygen and Radiation

Atomic oxygen (AO) is a highly reactive form of oxygen that is found in LEO. It can cause erosion and degradation of various materials on spacecraft, including polymers and lubricants, which can degrade their performance and durability [12,13] . At LEO altitudes, the neutral atmosphere primarily consists of AO (80%) and nitrogen molecules (20%). AO, which is formed by photodissociation of molecular oxygen in the upper atmosphere, is a severe hazard for material integrity [12]. A nominal range of values for atomic oxygen in LEO is $10^{16}$-$10^{17}$ atoms/m$^3$[12] .AO is highly reactive and readily oxidizes organic-based lubricants to form volatile products which can evaporate or sublimate. The effect can reduce lubricating film thickness and increase friction and wear of mechanisms [12]. AO can also oxidize metal-based lubricants such as Molybdenum Disulfide (MoS$_2$) and change surface properties, altering tribological performance [14]. Moreover, AO can interact synergistically with other environmental factors such as space debris and ultraviolet radiation and can increase the degradation of lubricants [14]. Therefore, it is important to select appropriate lubricants for LEO operations that can withstand effects of AO and maintain functionality.

Radiation in LEO can be categorized as ionizing or non-ionizing. Non-ionizing radiation in LEO includes the solar wind, radio waves, microwaves, and infrared radiation. The Van Allen radiation belts are a unique phenomenon in the Earth's atmosphere, consisting of two regions surrounding the Earth's magnetic field which is caused by both ionization and non-ionization radiation [11]. Ionization radiation in LEO is caused by various sources including trapped radiation in the Earth's radiation belts, galactic cosmic rays, and particles from solar events. These sources emit a range of particles such as high-energy electron flux, protons, alpha particles, heavy ions, and high-energy photons with energies ranging from several MeV to 100 MeV [15]. Electron flux, although penetrative, will not damage most oils and greases within lifetime of a satellite. However, some polymers such as Polytetrafluoroethylene (PTFE) may be affected in long term through increases in density and crystallinity [13].

## 3.2. Vacuum

Successful space devices rely on using lubricants with low evaporation rates, thermal stability, and radiation resistance [16]. These lubricants should be composed of properties such as thermal stability, low vapour pressure, and favourable viscosity & temperature properties. Specially, fluid lubricants are subjected to rapid evaporation where the rate of free evaporation of fluid under vacuum is given by the Langmuir equation [17].The LEO vacuum is typically between $10^{-9}$ Pa and $10^{-11}$ Pa and many fluids comprising even long chain molecules, used as lubricants in terrestrial applications, are too volatile for use in space. Volatile components characteristic of polymers are the low-molecular weight fragments, additives and absorbed gases [11]. Solid materials may also "outgas" resulting in the contamination of surfaces in a spacecraft, loss of dimensional stability as well as detrimental effects on material properties [5].

## 3.3. Temperature

In LEO, spacecraft experience time-dependent heat fluxes which produce thermal stress variations on the satellite and payload [18]. This requires a thermal control system specifically designed to mitigate thermal loads. There are also instances that lubricants may be challenged to operate in cyclic temperatures that can range from cryogenic conditions around 35 K up to 600 K [19] depending on the satellite position, orientation and exposure duration [6]. However, a rotational motion of a spacecraft can facilitate a more uniform thermal distribution across its surface [20]. The lowest temperatures found in space are significantly below the pour point of many space fluid lubricants, for instance the pour point of perfluorinated polyether (PFPE) fluid Fomblin Z25 is -75 °C (approx. 198 K) [21].

Conventionally, lubricant viscosity increases at low temperature, and this leads to greater power losses related to viscous shear. It has been demonstrated that torque values of angular contact bearings lubricated with space oils increase when temperature is decreased below 0°C, due to an increase in fluid viscosity that causes larger frictional losses [19]. Studies have shown that the lubricant YVAC40 increases the torque substantially, compared to other lubricants at lower temperature [16]. This outcome was echoed in the Space Mechanisms Lesson Learned Study, conducted in 1995 by NASA [22].

### 3.4. Microgravity

In LEO, the gravitational force experienced by an object is only marginally weaker than that on the Earth's surface [23]. This is due to the relatively small distance between LEO and the Earth's surface. However, an object in orbit experiences a state of weightlessness as it is in a continuous state of free fall around the Earth. This is a result of the balance between the gravitational force and the centrifugal force acting on the object [24]. This condition can alter the motion of liquid lubricants over solid surfaces as controlled by interfacial and capillary forces [24]. In addition, in a microgravity environment wear debris can contaminate or damage other components including mechanical parts and optics some distance away from their source [23].

### 4. TRIBOLOGICAL TESTS CLASSIFICATION

Tribological testing generally measures the friction and wear between two surfaces under defined loads, speeds and types of motion [25]. However, tribological tests can be performed by multitude of methods. The outcome of a tribotest is influenced by the characteristics of the material couple as well as the mechanical system and its environment. The process of selecting an appropriate test for a specific application is imperative in making meaningful prediction about the performance of a real mechanism [25].Tribological tests can be classified according to the degree of realism, i.e., how closely they imitate the conditions of a real application [25]. Generally, a high degree of realism is targeted. However, there are also many reasons to evaluate materials in tests which do not replicate an application [26]. DIB 50 322 is a German industrial standard proposed by Gahr in 1987 [27]. The standard categorizes tribo-tests according to their degree of realism, from the simplest material test, through model testing, simplified component testing, component testing, sub-system testing, bench testing and at the top-level field testing. In a model test, both tribo-surfaces are replaced by simulated components. A semi-tribocouple test is a simplified component test where only one surface is represented by a real component.

A full tribo-couple test is a component (or high-level) test where surfaces are represented by real components such as bearings and gears to evaluate their performance under more realistic settings [25]. Bench testing involves assessing the lubrication system components separately, or as a subsystem, on a test rig using simulated or real engine oil and conditions [25] . Field testing for lubrication involves evaluating the performance of lubricants in their intended operating environment. It involves testing the lubricants under real-world conditions to assess their effectiveness and suitability for their intended purpose. Figure 1 illustrates various forms of tribotest for a reaction wheel according to the degree of realism. Tribotests can also be classified as open and or closed [25], by consideration of the type of contact situation. If a surface continuously follows the same track on the counter body, the system is closed, whereas in an open test, the sliding track is continuously renewed [28].

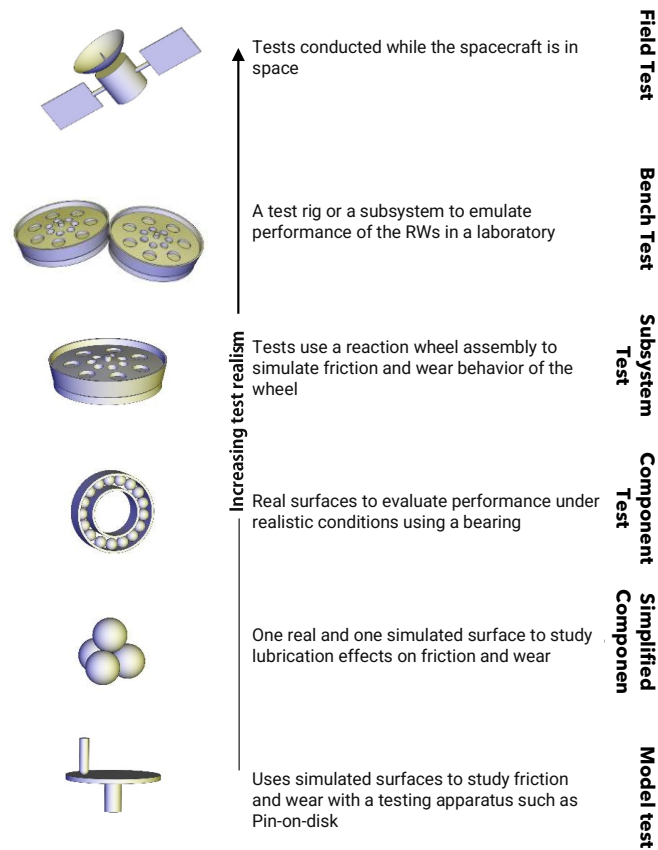

Figure 1. Classification of a reaction wheel tribo-test process according to degree of realism, adapted from Gahr (1987) and Bhushan (2000)

Tribotests with a high degree of realism can provide valuable insights into the degradation of lubrication, wear, and other characteristics of tribological mechanisms in space. However, conducting tests with a high degree of realism in space presents significant challenges due to the cost of space missions. In the next section, a thematic analysis of recent space-based tribotests categorized by degree of realism will be presented.

## 5. SPACE BASED TESTS

Over the past few years, several projects and space experiments have been dedicated to understanding the tribological characteristics of both materials and lubricants [14]. Initially, the primary means of investigating the space environment and its impact on materials and lubricants were limited to theoretical simulations and observations of material degradation on returning spacecraft [29].This work was augmented by ground based tribometer investigations, predominantly using an Eccentric Bearing Test Apparatus [30], Spiral orbit Tribometer (SOT) [31], Vacuum Four-Ball Tribometer [7], and Vacuum Pin-on-Disk Tribometer [6] Results from this various apparatus have been used to predict the performance of materials and lubricants in space. However, ground-based simulation facilities cannot replicate all the space environment factors accurately or simultaneously. Therefore, a number of space-based research projects have been conducted with varying degrees of realism in order to accurately comprehend the tribological behavior of space mechanisms. In the following sections, a thematic analysis of selected research is presented.

## 5.1. Model and simplified component tests

Testing materials for tribological uses remains at the forefront of materials development for space applications. One of the first extensive materials exposure experiments in space was NASA's Long Duration Exposure Facility (LDEF) [32]. LDEF was a cylindrical spacecraft that carried out 57 experiments to study the effects of outer space on various materials and systems. It was launched by the space shuttle Challenger in 1984 and retrieved by the space shuttle Columbia in 1990 [33]. It orbited Earth for more than five years and collected valuable data on the space environment. Since then, there have been several exposure experiments including experiments from the European Space Agency (ESA), Japan Aerospace Exploration Agency (JAXA) [34].

MISSE 7 is the seventh in a series of Materials International Space Station Experiments designed to provide a platform for conducting experiments in the harsh low earth orbit (LEO) environment [35]. The tribometers used in MISSE 7 measured friction and wear between different material pairs, including metals, ceramics, polymers, and composites. The experiment included over 32 passive exposure samples exposed to ultrahigh vacuum, temperature fluctuations, ultraviolet radiation, and atomic oxygen, as well as eight pin-on-disk tribometers. The pin-on-disk tribometer was designed and flown to the ISS for this project. The tribometers were delivered to the station in November 2009 and returned to Earth in April 2011. They had a normal force of 1 N, an inner radius of 1.6 mm, a wear track radius of 9 mm, a rotational disk speed of 14 RPM ± 0.05, and a linear sliding speed of 13.2 ± 0.1 mm/s. Various coatings were tested on the disk using the tribometer, including $MoS_2/Au/Sb_2O_3$, a continuous amorphous-nanocrystalline YSZ matrix surrounding nanoscopic inclusions of Au (YSZ/Au)/$MoS_2$/C, Polytetrafluoroethylene (PTFE)/nano-$Al_2O_3$, Gold, $MoS_2$/$Sb_2O_3$/Graphite, PTFE/nano-$Al_2O_3$, YSZ/Au/$MoS_2$/C and hydrogenated diamond-like carbon (DLC)/SiO-doped [35].

The Tribological Experiment Rig (TER) is another test set up with a pin-on-disk tribometer, which was designed to study behavior of wear debris in a microgravity environment. It was launched on board the Chinese manned spaceship Shenzhou-12 on 17th June 2021 [36]. The TER system consisted of illuminants, motors, a pin-on-disc tribometer, cameras, control panels, a power supply unit, a wear debris collecting device, and other supporting structures. The TER was launched and operated on orbit for 2 days and 19 hours and took 2,132 images of the wear debris. The images were analyzed and compared with the results obtained on earth under different gravity and medium (air) conditions. The experiment showed that the shape of the areas where the wear debris accumulated was different in microgravity than on earth. However, the motion path of the debris moved around the pin and the disk was the same in both cases. It was, therefore, concluded that gravity did not affect the motion path of the debris

## 5.2. Component and subsystem tests

Lubrication tests conducted in Spacelab 1 provided valuable insights into the motion of liquid lubricants over solid surfaces under the control of interfacial and capillary forces in a zero-gravity environment [24]. The tribology experiment consisted of two parts, which were performed with separate experimental hardware. The first part was a fluid wetting and spreading (FWS) study, and the second studied the morphology of a two-phase film under shearing between the surfaces utilizing journal bearing shaped devices. The investigation covered three journal bearing concepts including a plain journal bearing, a three-lobed bearing, and a plane bearing with locked-in eccentricity. The study provided a comprehensive understanding of fluid film movement and response under zero-gravity operating conditions. The system used photographs and vibration / displacement transducers to record data. Tribo-cosmos is a proposed experiment to test different materials and mechanisms on the ISS [37] within the Nauka module [38]. The experiment aims to use a new generation of pin-on-disk test apparatus with six stations and a bearing module to monitor and record data online. The experiment aims to study how friction, wear and lubrication are affected by space environmental factors such as vacuum, radiation, temperature, and wear debris. The experiment will use a new generation of tribo-test modules that can monitor and record data online. One test module is a pin-on-disc unit with 6 pins (3 on each side of disk) arranged at various radii [39]. The normal load is 0.1 N to 30 N, and the disk will have a rotational speed of 1000 rpm. The setup is intended to obtain online measurements of temperature, friction coefficient and wear. The other test module is comprised of a bearing arrangement subjected to a normal load of 10 N to100N, rotating at between 6 rpm and 1000 rpm. The overarching objective of the project is to compare the results of space and ground-based tribo-testing and to aid development of optimized conditions for simulating space environments.

## 5.3. Bench and field tests

In bench test, the tribological characteristics of mechanical components under simulated or actual operating conditions are considered. Generally, conducting a bench level test for a whole spacecraft is considered a challenging task. However, ground based tests conducted on the European Robot Arm (ERA) can be considered to be an example of a bench level test. The ERA is a robotic arm that connects to the Russian part of the ISS which was delivered and installed in 2021 [40,41] . It consists of two limbs, seven joints which provide the means for actuation. The ERA's Manipulator Joints Subsystem (MJS) underwent the main tests. Its bearings and gears were lubricated with Braycote 601 grease and lead ion-plated dry lubricants [40].Testing consisted of rotating the joints at various speeds and torques, applying dynamic and static braking operations, activating limit switches and hardtops, and measuring brake torque and joint stiffness. Several tests were performed in a vacuum chamber with a wide temperature range and a high vacuum level, to replicate the LEO environment [40].

Canadarm 2 is a 17-meter-long robotic arm on the International Space Station (ISS) and replaces the previous model named Canadarm 1 [42]. It is used to support station maintenance, moving supplies and equipment, and to catch and hold spacecraft. The arm has 6 degrees of freedom like a human arm and includes a shoulder (2 joints), elbow (1 joint) and wrists (3 joints). The robotic arm was designed to be refurbished in orbit (service missions were conducted in 2002, 2017 and 2018), and it provides a continuous tribological field test opportunity. The driving mechanism of the arm is a motor and gearbox, so the principal mechanisms are gears and rolling element bearings [42]. Molybdenum di-sulfide dry-film lubricant and a perfluoropolyether-based grease are used for lubrication and majority of the lubricants used are contained within housings and structures in such a way that mechanism is not exposed to AO. The deployment of other robotic arms in LEO [43,44] has enabled additional field level tribological tests. Given that the arms are exposed to the space environment and that LEO permits logistical convenience, they provide an ideal field-level test for determining tribological performance.

Solar array driving mechanisms are another component exposed to LEO environment which have been a focus of several tribological research [45]. The photovoltaic solar panels on the ISS track the sun via a constant rotating motion allow bearings in the Solar Array Alpha Rotary Joints on the main truss (SARJs). After installation in 2007, the port SARJ experienced increased torque and was closed-down due to exceeding driving current safety restrictions [46]. The failure of the SARJ mechanism was caused by several factors, some technical and some related to the development process. The mechanism was not adequately lubricated, which resulted in high friction and damage to the roller-race contact. The design was also vulnerable to roller tipping, which increased the surface forces and stresses. Finally, the mechanism was not fully tested, either analytically or experimentally, before being launched. On the positive side, the investigation highlighted that rolling contact with oils and greases is better at reducing friction than traditional solid lubricants with sliding contacts [46]. However, in the case studies of both Canadarm 2 and SARJ, the tests were conducted as post-study scenarios rather than dedicated field tests. As a result, there were many uncontrolled variables that could have introduced uncertainties in the outcomes of the studies.

## 6. DISCUSSION

Utilizing ground-based testing facilities is one of the primary strategies for space tribology research. Accelerated vacuum tribometers can emulate some of the space environment factors, such as vacuum, and temperature variation. However, ground based tribometers have limitations, for instance the difficulty of reproducing the effects of radiation flux and zero gravity. Moreover, these apparatuses tend to use simplified or model tests that do not reflect the complexity of real space mechanisms and lubrication complications. Therefore, there is a need for more realistic and comprehensive space-based platforms, such as MISSE 7 and Tribo-cosmos, that can expose materials and mechanisms to the real space environment and monitor their tribological behavior. These platforms have the advantage of providing direct and reliable data on the effects of space environment factors on friction, wear, and lubrication. However, even these platforms are not without drawbacks, such as the high cost, lengthy test design to mission launch time, and a low frequency of experiments. The use of space borne parts such as robotic arms and solar array driving mechanisms that can perform tasks and operations in space environments has opened new field level test opportunities. Robotic arms have the benefit of providing valuable insights into the performance and reliability mechanisms in space environments. However, the tests are prone to limitations, such as the difficulty of isolating and identifying the effects of specific variables on friction, wear, and lubrication. Moreover, robotic arms often use lubrication systems that are shielded within housings and structures that prevent direct observation and measurement of tribological behavior or direct exposure to the full space environment.

## 7. CONCLUSION

Space tribology is a challenging and important field of research that aims to understand and improve the performance and reliability of materials and mechanisms in space environments. The unique challenges presented by the LEO environment, including the presence of atomic oxygen, radiation, vacuum, temperature, and microgravity, necessitate extensive testing of space materials, mechanisms, and components to ensure continuous function and accuracy. Various experiments have been conducted in the past decades using different methods and technologies. This paper has presented a thematic review of tribology experiments conducted in Low Earth Orbit with different degrees of realism in the test scenario. The thematic analyses have ranged from model testing to full field testing and attempting to interpret the outcomes has highlighted the limitations and challenges of the collected data.